\documentclass[ twocolumn,aps,prd,   
               preprintnumbers,numbers,sort&compress,nofootinbib,
                            showpacs,
               colorlinks,
               linkcolor=blue,
               citecolor=blue]{revtex4-1}
   \newcommand{\exclude}[1]{}

\usepackage{graphicx,amsmath,amssymb,bm}
\usepackage{psfrag}
\usepackage{feynmp}
\usepackage{hyperref}
\usepackage{enumitem}

\newcommand{\beq}{\begin{equation}}
\newcommand{\eeq}{\end{equation}}
\newcommand{\be}{\begin{eqnarray}}
\newcommand{\ee}{\end{eqnarray}}

\def\+{\dagger}
\def\la{\langle}
\def\ra{\rangle}
\def\<{\langle}
\def\>{\rangle}

\newcommand{\Lqcd}{\Lambda_{\mathrm{QCD}}}

\begin{document}

\title{ ${\cal P}$ odd fluctuations  and Long Range Order in Heavy Ion Collisions.\\ Deformed QCD as   a Toy Model.}

\author{   Ariel R. Zhitnitsky} 
 \affiliation{Department of Physics \& Astronomy, University of British Columbia, Vancouver, B.C. V6T 1Z1, Canada}

\begin{abstract}
We study the local violation of  ${\cal P}$ and ${\cal CP}$ invariance in heavy ion collisions as observed at RHIC and LHC 
using a simple ``deformed QCD'' model. This model is a weakly coupled gauge theory, which however has all the relevant  crucial elements allowing us to study difficult and nontrivial questions which are known to be present in real strongly coupled QCD. Essentially, we want to understand the physics of   long range order in form of coherent  low dimensional   vacuum configurations observed in Monte Carlo lattice simulations.   Apparently precisely such kind of  configurations are  responsible for  sufficiently 
strong  intensity of asymmetries observed  in heavy ion collisions. 
 \end{abstract}

\maketitle

\section{Introduction and motivation} \label{introduction}
Recently it has become clear that quantum anomalies   play  very important role in the macroscopic dynamics of relativistic fluids. Much of this progress is motivated by very interesting ongoing  experiments 
on local ${\cal{P}}$  and ${\cal{CP}}$ violation in QCD as studied  at RHIC and ALICE at the LHC~ \cite{Voloshin:2004vk,Abelev:2009tx,collaboration:2011sma,Selyuzhenkov:2011xq,Abelev:2012pa}. It is likely that the observed asymmetry is due to   charge separation effect \cite{Kharzeev:2004ey,Kharzeev:2007tn} as a result of the chiral anomaly,  though some background processes may also   contribute to the measured observed asymmetry \cite{Bzdak:2009fc,Liao:2010nv,Bzdak:2010fd,Schlichting:2010qia,Pratt:2010zn}.
  The ideas formulated in \cite{Kharzeev:2004ey,Kharzeev:2007tn}  were further developed  in follow up papers \cite{Kharzeev:2007jp,Kharzeev:2007jp,Fukushima:2008xe}  where the effect was coined as chiral magnetic effect (CME). 

We  shall not discuss a number of subtle questions  of the  CME in the present work by  referring  to a recent review \cite{Kharzeev:2011vv}. Instead, we concentrate on a single  crucial element for CME to be operational. Namely, the key assumption of the proposal \cite{Kharzeev:2004ey,Kharzeev:2007tn}  is that the region where  the so-called $\la\theta (\vec{x}, t)_{ind}\ra\neq 0$  should be much larger in size than the scale of  conventional QCD fluctuations which have  typical     correlation lengths of order $\sim \Lambda_{QCD}^{-1}$. The  $\theta (\vec{x}, t)_{ind}$ parameter enters the effective lagrangian as follows,  ${\cal L_{\theta}}=-\theta_{ind} q$ where $ q \equiv \frac{g^2}{64\pi^2} \epsilon_{\mu\nu\rho\sigma} G^{a\mu\nu} G^{a\rho\sigma}$ is the topological density operator, such that local ${\cal{P}}$  and ${\cal{CP}}$
  invariance of QCD is broken on the scales where correlated state with $\la\theta (\vec{x}, t)_{ind}\ra\neq 0$  is induced. As a result of this violation, one should expect a number of ${\cal{P}}$  and ${\cal{CP}}$ violating effects taking place in a relatively large  region where $\la\theta (\vec{x}, t)_{ind}\ra\neq 0$.  The question which is addressed in the present work is as follows: what is the physics behind of this long range order when $\la\theta (\vec{x}, t)_{ind}\ra$ is correlated on distances much larger than conventional $\Lambda_{QCD}^{-1}$ scale?   

One should say that such kind of long range order indeed has been observed in recent lattice simulations which  have revealed some very unusual features.
 To be more specific, the gauge configurations studied in \cite{Horvath:2003yj,Horvath:2005rv,Horvath:2005cv,Alexandru:2005bn} display a laminar structure in the vacuum consisting of extended, thin, coherent, locally low-dimensional sheets of topological charge embedded in 4d space, with opposite sign sheets interleaved. A similar structure has been also 
observed in QCD by  different groups~\cite{Ilgenfritz:2007xu,Ilgenfritz:2008ia,Bruckmann:2011ve,Kovalenko:2004xm,Buividovich:2011cv,Blum:2009fh} and also in two dimensional $CP^{N-1}$ model~\cite{Ahmad:2005dr}.  

The key  element relevant for the present studies is the observation that the important gauge configurations which contribute to the CME are in fact  very {\it similar} to extended, locally low-dimensional sheets of topological charge embedded in 4d space studied in refs.\cite{Horvath:2003yj,Horvath:2005rv,Horvath:2005cv,Alexandru:2005bn,Ilgenfritz:2007xu,Ilgenfritz:2008ia,Bruckmann:2011ve,Kovalenko:2004xm,Buividovich:2011cv,Blum:2009fh}.
Such a similarity was   noticed  previously using numerical lattice simulations  \cite{Buividovich:2011cv,Blum:2009fh}. The main goal of the present work is to understand this relation on a deeper theoretical level using ``deformed QCD'' as a toy model.  Furthermore, we shall argue  that  an external  magnetic field (which is a key element in  studying CME)
may serve as  a complimentary tool to analyze the relevant topological configurations present in the QCD vacuum. 

We start in section~\ref{deformedqcd}  by reviewing  the relevant parts of the model \cite{Yaffe:2008,Thomas:2011ee}. This is a simplified  version of QCD which, on one hand, is  a weakly coupled gauge theory wherein computations can be performed in theoretically controllable manner. On other hand, the corresponding  deformation  preserves all the relevant elements of strongly coupled QCD such as confinement, degeneracy of topological sectors, nontrivial $\theta$ dependence, and many other important  aspects which allow us to test some fascinating   features of strongly interacting QCD, including the CME and long range order as we shall argue below. 

In section \ref{DW_eta} we introduce the matter field into the model such that it   couples to the  external $U(1)$ magnetic field. We observe that  the matter field essentially traces the long range order which is inevitable feature of  pure gauge theory. 

In section {\ref{CME} we apply these ideas to CME. We  observe that CME  takes place precisely on the same extended  configurations  which  are identified with  low-dimensional sheets   studied in lattice simulations \cite{Horvath:2003yj,Horvath:2005rv,Horvath:2005cv,Alexandru:2005bn,Ilgenfritz:2007xu,Ilgenfritz:2008ia,Bruckmann:2011ve,Kovalenko:2004xm,Buividovich:2011cv,Blum:2009fh}. In different words, the CME which takes place  in the presence of the magnetic field precisely {\it traces} the structure of the original topological configurations which are always present in QCD vacuum. 
This observation suggests that an external magnetic field could be a   perfect complimentary tool which allows to study a very complicated QCD vacuum structure which apparently shows a long range order rather than some localized finite size fluctuations such as instantons.

 Section \ref{conclusion} is our conclusion where we   consider applications of these ideas to  ${\cal{P}}$ odd fluctuations 
 observed  at  RHIC and ALICE at the LHC~ \cite{Voloshin:2004vk,Abelev:2009tx,collaboration:2011sma,Selyuzhenkov:2011xq,Abelev:2012pa}. We interpret  
  a relatively strong asymmetry observed in heavy ion collisions as a  direct manifestation of the long range order studied in present work. We argue that the asymmetry would be much weaker than observed if this long range order was  not present in the system.

\section{Deformed QCD} \label{deformedqcd}

Here we overview  the ``center-stablized" deformed Yang-Mills developed in \cite{Yaffe:2008} and reviewed in \ref{model}. In the deformed theory an extra term is put into the Lagrangian in order to prevent the center symmetry breaking that characterizes the QCD phase transition between ``confined" hadronic matter and ``deconfined" quark-gluon plasma. Thus we have a theory which remains confined at high temperature in a weak coupling regime, and for which it is claimed \cite{Yaffe:2008} that there does not exist an order parameter to differentiate the low temperature (non-abelian) confined regime from the high temperature (abelian) confined regime.  One should remark here that  ``deformed QCD'' model has been previously successfully  used to test some  highly nontrivial features of strongly coupled QCD such as emergence of non-dispersive contact term in topological susceptibility \cite{Thomas:2011ee} and emergence of the topological Casimir   behaviour in gauge theory  with a gap \cite{Thomas:2012ib}. Finally, what is most relevant for present studies is the observation that ``deformed QCD'' model exhibits a long range order and double layer structure which was interpreted in ref. \cite{Thomas:2012tu} as a manifestation of the corresponding  features observed in strongly coupled regime in the lattice   simulations. 
The extended structure in this model is manifested in a form of the domain walls, which are present in the system as  a result of a generic $2\pi$ periodicity of the effective low energy Lagrangian   governing  the dynamics of ``deformed QCD".  
The low energy description will be reviewed in section \ref{lagrangian} while the part  on domain walls in pure glue theory will be reviewed in section \ref{dw}.

\subsection{The model}\label{model}

We start with pure Yang-Mills (gluodynamics) with gauge group $SU(N)$ on the manifold $\mathbb{R}^{3} \times S^{1}$ with the standard action
\be \label{standardYM}
	S^{YM} = \int_{\mathbb{R}^{3} \times S^{1}} d^{4}x\; \frac{1}{2 g^2} \mathrm{tr} \left[ F_{\mu\nu}^{2} (x) \right],
\ee
and add to it a deformation action,
\be \label{deformation}
	\Delta S \equiv \int_{\mathbb{R}^{3}}d^{3}x \; \frac{1}{L^{3}} P \left[ \Omega(\mathbf{x}) \right],
\ee 
built out of the Wilson loop (Polyakov loop) wrapping the compact dimension
\be \label{loop}
	\Omega(\mathbf{x}) \equiv \mathcal{P} \left[ e^{i \oint dx_{4} \; A_{4} (\mathbf{x},x_{4})} \right].
\ee
Parameter  $L$ here  is the length of the compactified dimension
which is assumed to be small. 
 The coefficients of the polynomial  $ P \left[ \Omega(\mathbf{x}) \right]$ can be suitably chosen such that the deformation potential (\ref{deformation}) forces unbroken symmetry at any compactification scales. At small compactification $L$ the gauge coupling  is small so that 
the semiclassical computations are under complete theoretical control \cite{Yaffe:2008}.

\subsection{Infrared description}\label{lagrangian}

As described in \cite{Yaffe:2008}, the proper infrared description of the theory is a dilute gas of $N$ types of monopoles, characterized by their magnetic charges, which are proportional to the simple roots and affine root $\alpha_{a} \in \Delta_{\mathrm{aff}}$ of the Lie algebra for the gauge group $U(1)^{N}$. 
 For a fundamental monopole with magnetic charge $\alpha_{a} \in \Delta_{\mathrm{aff}}$, the topological charge is given by
\be \label{topologicalcharge}
	Q = \int_{\mathbb{R}^{3} \times S^{1}} d^{4}x \; \frac{1}{16 \pi^{2}} \mathrm{tr} \left[ F_{\mu\nu} \tilde{F}^{\mu\nu} \right]
		= \pm\frac{1}{N},
\ee
and the Yang-Mills action is given by
\be \label{YMaction}
	S_{YM} = \int_{\mathbb{R}^{3} \times S^{1}} d^{4}x \; \frac{1}{2 g^{2}} \mathrm{tr} \left[ F_{\mu\nu}^{2} \right]= \frac{8 \pi^{2}}{g^{2}} \left| Q \right|.
		 \ee
 The  $\theta$-parameter in the Yang-Mills action can be included in conventional way,
\be \label{thetaincluded}
	S_{\mathrm{YM}} \rightarrow S_{\mathrm{YM}} + i \theta \int_{\mathbb{R}^{3} \times S^{1}} d^{4}x\frac{1}{16 \pi^{2}} \mathrm{tr}
		\left[ F_{\mu\nu} \tilde{F}^{\mu\nu} \right],
\ee
with $\tilde{F}^{\mu\nu} \equiv \epsilon^{\mu\nu\rho\sigma} F_{\rho\sigma}$.

The system of interacting monopoles, including $\theta$ parameter, can be represented in the dual sine-Gordon form as follows \cite{Yaffe:2008,Thomas:2011ee},
\be
\label{thetaaction}
	S_{\mathrm{dual}}&=& \int_{\mathbb{R}^{3}}  d^{3}x \frac{1}{2 L} \left( \frac{g}{2 \pi} \right)^{2}
		\left( \nabla \bm{\sigma} \right)^{2} \nonumber \\&-& \zeta  \int_{\mathbb{R}^{3}}  d^{3}x \sum_{a = 1}^{N} \cos \left( \alpha_{a} \cdot \bm{\sigma}
		+ \frac{\theta}{N} \right)  	, 
\ee
where $\zeta$ is magnetic monopole fugacity which can be explicitly computed in this model using conventional semiclassical approximation. The $\theta$ parameter enters the effective Lagrangian (\ref{thetaaction}) as $\theta/N$ which is the direct consequence of the fractional topological charges of the monopoles (\ref{topologicalcharge}). Nevertheless, the 
theory is still $2\pi$ periodic. This
  $2\pi$ periodicity of the theory is restored not due to the $2\pi$ periodicity of Lagrangian (\ref{thetaaction}).
  Rather, it is restored as a result of   summation over all branches of the theory when the  levels cross at
   $\theta=\pi (mod ~2\pi)$ and one branch replaces another and becomes the lowest energy state as discussed in \cite{Thomas:2011ee}.
 
Finally, the dimensional parameter which governs the dynamics  of the problem is the Debye   correlation length of the monopole's gas, 
 \be \label{sigmamass}
	m_{\sigma}^{2} \equiv L \zeta \left( \frac{4\pi}{g} \right)^{2}.
\ee
   The average number of monopoles in a ``Debye volume" is given by
\begin{equation} \label{debye}
{\cal{N}}\equiv	m_{\sigma}^{-3} \zeta = \left( \frac{g}{4\pi} \right)^{3} \frac{1}{\sqrt{L^3 \zeta}} \gg 1,
\end{equation} 
The last inequality holds since the monopole fugacity is exponentially suppressed, $\zeta \sim e^{-1/g^2}$, and in fact we can view (\ref{debye}) as a constraint on the validity of the  approximation where semiclassical approximation is justified. 
 
 \subsection{Domain Walls  in deformed QCD} \label{dw}
A  discrete set of degenerate vacuum states as a result of the  $2\pi $ periodicity of the effective Lagrangian  (\ref{thetaaction}) for $\bm{\sigma} $ field  is a signal that the domain  wall configurations 
interpolating between these   states are present in the system.  However,  the corresponding configurations are not conventional static domain walls similar to the well known ferromagnetic domain walls in condensed matter physics which interpolate between physically {\it distinct} vacuum states. In contrast, in present case a corresponding  configuration interpolates between 
      topologically different but physically equivalent  winding states $| n\ra$, which are connected  to each other by large gauge transformation operator.   Just because of that,     the corresponding domain wall configurations  in Euclidean space should be  interpreted  as configurations describing the tunnelling processes in Minkowski space, similar to  Euclidean monopoles which also interpolate 
      between topologically different, but physically identical states, see detail discussions  in  \cite{Thomas:2012tu} 
      and references therein. 
      
      One should remark that  a formal similar construction  has been considered previously in  relation with the  so-called $N=1$ axion model \cite{Vilenkin:1982ks,Huang:1985tt,Hagmann:2000ja}, and more recently in QCD context in \cite{Forbes:2000et} and in high density QCD in \cite{Son:2000fh}. In  previously considered cases \cite{Forbes:2000et,Son:2000fh} as well as in recently considered  case \cite{Thomas:2012tu}   there is a single physical unique vacuum state, and interpolation    corresponds to the transition from one to the same physical state.

 There are $N$ different domain wall (DW) types. However,  there are only $(N-1)$ physical propagating scalars  $\bm{\sigma}$ in the system as one singlet scalar  field, though it remains massless,   completely decouples from the system, and does not interact with other components at all \cite{Yaffe:2008}.  
 
 In what follows, without loosing any generalities,  we consider    $N=2$ case. In this case there is only  one physical field $\chi$ which corresponds to a single diagonal component from the original $SU(2)$ gauge group.  The  action (\ref{thetaaction}) becomes, 
  \be
 \label{action}
S_{\chi}&=& \int_{\mathbb{R}^{3}}  d^{3}x \frac{1}{4 L} \left( \frac{g}{2 \pi} \right)^{2}
		\left( \nabla \chi \right)^{2} \\ &-&    \zeta  \int_{\mathbb{R}^{3}}  d^{3}x  \left[\cos \left( \chi 
		+ \frac{\theta}{2} \right) +  \cos \left(- \chi 
		+ \frac{\theta}{2} \right) \right], \nonumber
		\ee
		while the equation of motion and its solution  
 take   the form~ \cite{Thomas:2012tu}
	\be
 \label{solution}		
\nabla ^2 \chi &-&m_{\sigma}^2\sin\chi=0, \nonumber\\
   \chi (z)&=& 4\arctan \left[\exp(m_{\sigma}z)\right]
 \ee
 where we  take $\theta=0$ to simplify things.
  The width of the domain wall is obviously determined by $m_{\sigma}^{-1}$, while the domain wall tension $\sigma$ for profile (\ref{solution}) can be easily computed and it is given by
 \be
 \label{tension}
 \sigma&=& 2\cdot \int^{+\infty}_{-\infty}   dz \frac{1}{4 L^2} \left( \frac{g}{2 \pi} \right)^{2}
		\left( \nabla\chi \right)^{2}\nonumber\\ &=&  \frac{m_{\sigma}}{L^2} \left( \frac{g}{2 \pi} \right)^{2} \sim  \sqrt{\frac{\zeta}{L^3}}.
 \ee
    The topological charge density for profile (\ref{solution}) assumes a double layer structure as discussed in details in  \cite{Thomas:2012tu}
 \be
 \label{Q}
 q(z)=\frac{\zeta}{L}\sin\chi (z)=\frac{4\zeta}{L} \frac{ (e^{m_{\sigma}z}-e^{-m_{\sigma}z}) }{(e^{m_{\sigma}z}+e^{-m_{\sigma}z})^2}.
 \ee
 From eq. (\ref{Q}) one can explicitly see that the net topological charge $Q\sim\int^{\infty}_{-\infty} dz q(z)$ on the domain wall obviously vanishes. However, the charge density is distributed not uniformly. Rather, it is organized in a double layer structure. It   was conjectured in   \cite{Thomas:2012tu} that this structure is in fact  a trace of a similar configurations  studied 
  in the  lattice simulations. 
 
  One should also comment that, formally, a similar soliton -like solution which follows from  action (\ref{action}) appears in  computation of the string tension in 3d Polyakov's model \cite{Polyakov,Yaffe:2008}.  The solution considered there emerges  as a result of insertion of the external sources in a course of computation of the vacuum expectation of the Wilson loop. In contrast, in our case, the solution (\ref{solution}) is internal part of the system without any external sources. 
 Furthermore, the physical meaning of these solutions are fundamentally different:  in our case the interpretation of solution (\ref{solution}) is  similar to instanton describing the tunnelling processes in Minkowski space, while in  computations  \cite{Polyakov,Yaffe:2008} it was an auxiliary object which appears in the course of computation of the string tension.

As we already mentioned the DW described here are classically stable objects, but quantum mechanically they decay as a result of tunnelling processes. The corresponding rate was estimated in  \cite{Thomas:2012tu} and is given by
                   \be
         \label{rate}
         \Gamma &\sim&  
          \exp \left(- \pi^3\left(\frac{g}{4\pi}\right)^3\frac{ \ln^2  (\frac{1}{m_{\sigma}L} )}{\sqrt{L^3\zeta}}\right) \nonumber\\
         &\sim& \exp \left(-\gamma\cdot {\cal{N}}\ln^2 {\cal{N}} \right) \ll 1,
         \ee
         with $\gamma$ being some numerical coefficient and ${\cal{N}}\gg 1 $  is   large parameter of the model (\ref{debye}).
           The estimate (\ref{rate}) justifies our treatment of   the domain walls as the 
             stable objects  when the weak coupling regime is enforced  by the deformation (\ref{deformation}). 
                  
 We conclude this short overview   by emphasizing that the most important lesson  from  analysis   \cite{Thomas:2012tu} is  that the double layer structure represented by eq. (\ref{Q}) naturally emerges in construction of the domain walls in weak coupling regime in deformed QCD.
 As claimed in \cite{Yaffe:2008} the transition from  high temperature weak coupling regime to low temperature   strong coupling  regime  should be smooth without any phase transitions on the way. 
 As a consequence of this smoothness,     the double layer structure   (\ref{Q})  was interpreted in  \cite{Thomas:2012tu} as a trace  of  the double layer structure observed in the  lattice simulations   in strong coupling regime  \cite{Horvath:2003yj,Horvath:2005rv,Horvath:2005cv,Alexandru:2005bn}. In different words, as
  the transition from   weak coupling regime to    strong coupling  regime  should be smooth in this model it is naturally to assume that
  the domain walls (\ref{solution})  become very clumpy with large number of folders. Such fluctuations obviously increase the entropy of the DW which eventually may overcome the intrinsic tension (\ref{tension}).  If this happens, the DWs would look like as very crumpled and wrinkled  objects with large number of foldings and effectively vanishing tension as a result of large entropy. In this case an arbitrary number of such objects can be formed and they can percolate through the vacuum, forming a kind of a vacuum condensate.  Furthermore,  the DWs may loose their natural dimensionality, and likely to  be characterized by a Hausdorff dimension  as recent lattice simulations suggest  \cite{Buividovich:2011cv}. Nevertheless, the  topological charge distribution with  such  striking  features  as extended coherence along $x,y$ directions   and   double  layer structuring along $z$ direction, see eq.(\ref{Q}),   should   still  persist     as the transition  from weak to strong   coupling regime should be smooth. 
  It is quite likely that an appropriate description for this physics should be formulated in terms of holographic dual model as argued in \cite{Zhitnitsky:2011aa}, however we leave this subject for future studies.

  \section{Domain Walls  in the presence of  matter field}\label{DW_eta}
  The  ultimate goal of the present work is to understand  the long range structure described above in section \ref{dw} in the presence of external magnetic field, which is precisely the environment relevant for study of the CME. However, the gluons in pure glue theory represented in this model by low energy effective Lagrangian (\ref{thetaaction}) do not couple to physical magnetic field. Therefore, we introduce    a single charged massless quark field $\psi$ into the model.  The low energy description of the system in confined phase with a single quark is accomplished by introducing the $\eta'$ colour singlet, charge neutral, meson. As usual, the $\eta'$ would be conventional massless Goldstone boson if the chiral anomaly is ignored. In the dual sine-Gordon theory   the $\eta'$ field   appears  exclusively in combination with the $\theta$ parameter as $\theta \rightarrow \theta - \eta'$ as a consequence of the Ward Identities. Indeed, the transformation properties of the path integral measure under the chiral transformations $\psi\rightarrow \exp(i \gamma_5\frac{\eta'}{2})\psi$ dictate  that $\eta'$ appears only in the combination  $\theta \rightarrow \theta - \eta'$.  Therefore we have,
\be
\label{matter}
	S_{\mathrm{dual}}&=& \int_{\mathbb{R}^{3}}  d^{3}x \frac{1}{2 L} \left( \frac{g}{2 \pi} \right)^{2}
		\left[ \left( \nabla \bm{\sigma} \right)^{2} + \frac{c}{2} \left( \nabla \eta' \right)^{2}\right] \nonumber \\&-& \zeta  \int_{\mathbb{R}^{3}}  d^{3}x \sum_{a = 1}^{N} \cos \left( \alpha_{a} \cdot \bm{\sigma}
		+ \frac{\theta-\eta'}{N} \right)  	, 
\ee
where dimensionless numerical coefficient $c\sim 1$  can be, in principle, computed in this model, though our results do not depend on its numerical value.  
One can explicitly compute the topological susceptibility $\chi$ in this model, and check that the Ward Identities are automatically satisfied when the $\eta'$ field enters the Lagrangian precisely in form (\ref{matter}), see the detail computations in \cite{Thomas:2011ee}. The $\eta'$ mass computed from (\ref{matter}) has an extra  $1/N$ suppression, as it should, 
 in comparison with mass of the $ \bm{\sigma}$ field
\be
\label{mass}
m_{\eta'}^{2} = \frac{2L \zeta}{cN} \left( \frac{2\pi}{g} \right)^2, ~~~~ \frac{m_{\eta'}^2}{m_{\sigma}^2}=\frac{1}{2cN}.
\ee

As we already mentioned, the system (\ref{matter}) is $2\pi$ periodic  with respect $\theta\rightarrow \theta+2\pi$. 
This $2\pi$  periodicity of the theory  is not explicit in eq. (\ref{matter}), but nevertheless is restored as a result of summation over all branches of the theory   as explained  in  \cite{Thomas:2011ee}. The fact that $\eta'$ field enters the low energy effective Lagrangian precisely in combination  $ (\theta-\eta') $ implies that  system (\ref{matter}) is also periodic with respect to shift $\eta'\rightarrow \eta'+2\pi$ when summation over all branches of the theory is properly implemented. In different words, the system (\ref{matter})  supports $\eta' (z) $ domain walls as a result of a very  generic feature of the theory\footnote{There are many different types of the domain walls which are supported by the Lagrangian (\ref{matter}). We leave this problem of classification of the DWs for a future study. In this work we concentrate on a simplest possible case with $N=2$ to demonstrate few generic features of the system.}. This argument has been previously used to construct the $\eta'$ domain walls in context  of the axion physics \cite{Forbes:2000et}. One can derive the equation of motion for set of $\eta'(z) $ and $ \bm{\sigma} (z) $ fields determined by  Lagrangian (\ref{matter}) and   impose an appropriate boundary conditions $\eta' (z=+\infty)-\eta' (z=-\infty)=2\pi$ to  analyze this system numerically, in close analogy with  procedure used, though in a different context,  in ref.\cite{Forbes:2000et}. 

However, for our present work it  is sufficient to qualitatively describe the behaviour of  the system in the limit when $ {m_{\eta'}^2}/{m_{\sigma}^2}\ll 1$, when the  basic features can be easily understood even without numerical computations. The most important  lesson of this qualitative analysis,  as we shall see in a moment,  is that the light $\eta'$ field traces  the $ \bm{\sigma} (z) $ field by exhibiting  a similar  double layer structure    discussed for pure gauge theory  in section \ref{dw}.

The low energy Lagrangian which describes the lightest degrees of freedom for $SU(2)$ gauge group is governed by the following action 
  \be
 \label{chi_eta}
&S_{ \eta'}&= \int_{\mathbb{R}^{3}}  d^{3}x \frac{1}{4 L} \left( \frac{g}{2 \pi} \right)^{2}
		\left[\left( \nabla \chi \right)^{2} +c \left( \nabla \eta' \right)^{2}\right]  \\ &-&    \zeta  \int_{\mathbb{R}^{3}}  d^{3}x  \left[\cos \left( \chi 
		+ \frac{\theta-\eta'}{2} \right) +  \cos \left(- \chi 
		+ \frac{\theta-\eta'}{2} \right) \right], \nonumber
		\ee
		where we   inserted the $\eta'$ field into eq. (\ref{action}) exactly in the form consistent with Ward Identities.  
Now,  it is convenient to represent the action (\ref{chi_eta}) of the system in the following way
  \be
 \label{eta}
S_{ \eta'}&=& \int_{\mathbb{R}^{3}}  d^{3}x \frac{1}{4 L} \left( \frac{g}{2 \pi} \right)^{2}
		\left[\left( \nabla \chi \right)^{2} +c \left( \nabla \eta' \right)^{2}\right]  \\ &-&    2\zeta  \int_{\mathbb{R}^{3}}  d^{3}x  \left[\cos  \chi 
		\cdot\cos \left(\frac{\eta'}{2} \right) \right],  \nonumber
		\ee
where we take $\theta=0$ to simplify things. We are looking for  a DW solution 
which satisfies the following boundary conditions:
\be
\label{bc}
(\chi&\rightarrow& 0, \eta'\rightarrow 0)~~~ {\rm as} ~~~z \rightarrow -\infty \\
(\chi &\rightarrow &\pi, \eta'\rightarrow 2\pi)~~~ {\rm as} ~~~z \rightarrow +\infty,   \nonumber
\ee
 One can explicitly see from (\ref{eta}) that the vacuum energy for $(\chi =\pi, \eta'= 2\pi)$ at $z=+\infty$ 
is identically coincide with vacuum energy when $(\chi, \eta')$ fields assume their  trivial vacuum values: $(\chi= 0, \eta'= 0)$ at  $z=-\infty$.  
 
 As we already  
emphasized these states (with boundary conditions $\{\chi= 0, \eta'= 0\}$ and $\{\chi =\pi, \eta'= 2\pi\}$ correspondingly) must be interpreted as topologically different but physically equivalent states.  Therefore, the corresponding domain wall configurations in Euclidean space should be interpreted as configurations describing the tunnelling processes rather than real DW in Minkowski space-time.
As we discussed   in \cite{Thomas:2012tu}, a similar domain wall which has an analogous interpretation is known to exist in QCD at large temperature in weak coupling regime where it can be described in terms of classical equation of motion.
These are so-called $Z_N$ domain walls which separate domains characterized by a different value for the Polyakov loop at high temperature.
As is known, see the review papers \cite{smilga,Fukushima:2011jc} and references therein, these $Z_N$ domain walls interpolate between topologically different but physically identical states connected by large gauge transformations similar to our case. 

Coming back to our case, the boundary conditions (\ref{bc}) correspond to the case when the large gauge transformation represented (in this model) by the   $\chi$ field is compensated by the $\eta'$ field which couples with the gluon density operator (\ref{thetaincluded}) as $\eta'$ enters the action in combination $(\theta-\eta')$ as explained above. 
The corresponding equations of motion in this simple $N=2$ case
take the form
\be
\label{set}
\frac{1}{m_{\sigma}^2}\nabla ^2 \chi &=&\sin\chi \cdot \cos \left(\frac{\eta'}{2}\right),\\
\frac{1}{2m_{\eta'}^2}\nabla ^2 \eta'&=& \cos \chi\cdot\sin\left(\frac{\eta'}{2}\right).\nonumber
\ee
We are looking for a  solution of the system (\ref{set}) which satisfies the boundary conditions (\ref{bc}).

It is interesting to note that  our system (\ref{set}) with boundary conditions (\ref{bc}) formally is identical to  the  axion DW configuration
analysed  in ref. \cite{Huang:1985tt} in the limit when the  isotopical symmetry is exact, i.e. $m_u=m_d$.
In this case our $\chi$ field plays the role of the  $\pi^0$ field denoted as $\gamma$ in ref.\cite{Huang:1985tt} while $\frac{\eta'}{2}$ plays the role of  the axion field $\alpha$ from \cite{Huang:1985tt}. Precise relations  are as follows: 
\be
\chi&\rightarrow &\pi-\gamma, ~~ m_{\sigma}^2\rightarrow m_{\pi}^2 \\
\frac{\eta'}{2}&\rightarrow & \alpha, ~~~ m_{\eta'}\rightarrow m_{a}^2
\label{analogy}
\ee
 such that our equations (\ref{set}) and boundary conditions (\ref{bc}) are identically coincide with the equations and the  boundary conditions studied  in ref \cite{Huang:1985tt}. Therefore, we simply formulate  here the main points\footnote{A simplest intuitive way to  understand
 the qualitative behaviour of the system (\ref{set}) is to use a mechanical analogy  as suggested in \cite{Huang:1985tt}  when variable $z$ is replaced by time, while the fields $(\eta',  \chi)$ 
 can be thought as coordinates of two particle moving in one dimension with interaction determined by the potential term from eq.(\ref{eta}).}
 which are relevant for our studies  referring for the technical details to ref.\cite{Huang:1985tt}.
 
  Most important result of analysis of ref.  \cite{Huang:1985tt} is that there is a unique solution of equations (\ref{set}) which satisfies  the boundary conditions (\ref{bc}). The simplest way to convince yourself that such a solution should exist is to use  a mechanical analogy  as suggested in \cite{Huang:1985tt}.
While analytical formulae of the solutions  is not known, its asymptotical behaviour at very large distances   $|z|\gg m_{\sigma}^{-1}$ can be easily found as follows.
It is clear that al large negative $z$  a heavy $\chi (z)$ field already assumes its vacuum value
 $\chi=0$ such that $\eta'(z)$ domain wall equation  can be approximated in this region as  
\be
 \label{eta2}
  \nabla ^2 \eta' -2m_{\eta'}^2\sin\left(\frac{\eta'}{2}\right)\simeq 0,~~   z\ll -\frac{1}{m_{\sigma}}.
 \ee
Solution of this equation which vanishes  at large distances $\eta'(z\rightarrow-\infty)\rightarrow 0$   can be approximated as 
\be
 \label{eta3}
 \eta'(z)=8 \arctan \left[\tan\left(\frac{\pi}{8}\right)e^{m_{\eta'} z}\right], ~~  z\ll -\frac{1}{m_{\sigma}}.
  \ee
  Similarly, 
    al large positive $z$  a heavy $\chi (z)$ field already assumes its vacuum value
 $\chi=\pi$ such that $\eta'(z)$ domain wall equation  can be approximated in this region as  
\be
 \label{eta4}
  \nabla ^2 \eta' +2m_{\eta'}^2\sin\left(\frac{\eta'}{2}\right)\simeq 0,~~   z\gg \frac{1}{m_{\sigma}}.
 \ee
Solution of this equation which approaches its vacuum value $\eta'=2\pi$  at large distances $\eta'(z\rightarrow+\infty)\rightarrow 2\pi$   can be approximated as 
\be
 \label{eta5}
 \eta'(z)=2\pi-8 \arctan \left[\tan\left(\frac{\pi}{8}\right)e^{-m_{\eta'} z}\right], ~z\gg \frac{1}{m_{\sigma}}.~~~
  \ee
  Formally, a similar construction when   $\eta' (z)$ field interpolates between different branches representing the same physical vacuum state was considered previously, see \cite{Forbes:2000et} for the details and earlier references on the subject\footnote{The crucial difference with \cite{Forbes:2000et}  is of course that the solutions  for the system of the fields $(\eta', \chi)$ considered here are regular functions everywhere, while solution in ref.  \cite{Forbes:2000et}  had a cusp singularity as a result of integrating out heavy fields played by the $\chi$  field in present ``deformed QCD" model. Interpretations of these solutions in these two cases  are  also very different as we interpret the corresponding  configurations as the transitions describing the tunnelling processes in Euclidean space-time rather than real static DW solution in Minkowski space-time as we mentioned above.}.

What happens to  the double layer structure (\ref{Q}) for the topological charge distribution in the presence of the light dynamical quark? We  anticipate, without any computations,  that  the light dynamical  field suppresses the topological fluctuations similar to analysis of the topological susceptibility in this model \cite{Thomas:2011ee}. Indeed, one can  
support this expectation by the following argument. As the first step we 
 represent $q(z)$ 
 in form similar to eq. (\ref{Q}). The only difference in comparison with previous formula   (\ref{Q}) is an emergence of the  extra term due to the $\eta'$ field    on the right hand side of eq. (\ref{Q_eta}),  
 \be
 \label{Q_eta}
 q(z)=\left(\frac{g}{4\pi}\right)^2\frac{\nabla^2 \chi }{L^2} =\frac{\zeta}{L}\sin\chi (z)\cos\left(\frac{\eta'(z)}{2}\right).
  \ee
  In obtaining  (\ref{Q_eta}) we used equations (\ref{set}) and (\ref{sigmamass}) to simplify  the expression for $q(z)$. 
  As we already mentioned, the $\chi(z)$ field has a solitonic shape interpolating between $\chi=0$ and $\chi=\pi$, see eq. (\ref{bc}). Therefore, $q(z)\sim \partial^2 \chi/\partial z^2$ inevitably produces a  double layer structure irrespectively to the details of  the solitonic shape of $\chi$, 
  similar to our discussions of a pure gauge theory in section \ref{dw}. However, the magnitude of
   this structure is  strongly suppressed as a result of dynamics of  the light quark. Indeed, the $\eta'(z)$ field assumes its central value $\eta'\approx \pi$ in the  region where $\chi (z)$ field   varies and $\partial^2 \chi/\partial z^2$ is sufficiently large.  This factor $\cos \frac{\eta'}{2} $ leads to a strong  suppression of $q(z)$ in eq.(\ref{Q_eta}) as anticipated. 
   
   Few comments are in order. 
First of all, if we would introduce  the quark in ``quenched approximation" rather than as a dynamical degree of freedom,   we would return to  our pure glue expression (\ref{Q}) as the ``quenched approximation"  corresponds to the $\eta'=0$ in formula (\ref{Q_eta}).
Secondly,  if we would introduce a small quark's mass $m_{\psi}\neq 0$ into our Lagrangian it would not change qualitative picture presented above as boundary conditions
imposed on the system (\ref{bc}) can not depend on $m_{\psi}\neq 0$. Indeed, these boundary conditions    are entirely  based on exact symmetry 
of the system which requires that the energy of the system  at $\theta=0$ and $\theta=2\pi$ is identically the same irrespectively to the quark's mass. 
 Finally,  the $(\chi, \eta')$ domain wall   is a coherent configuration and   can not easily decay into its constituents,   even though the $\eta'$ is a real physical asymptotic state of the system. This DW   can only decay through the tunnelling process  similar to our previous discussions, see eq. (\ref{rate}). Technically, it can be also explained using pure kinematical arguments: the $\eta'$ constituents which are making the $\eta'$ domain wall are   off-shell, rather than on-shell, states.  
 
 To conclude this section: the main lesson of the present analysis  is that   the deformed QCD with matter field supports long range correlated configurations.
In different words,  the matter and glue fields accompany each other in their interpolations between topologically different, but physically identical states. This correlation is enforced by very generic features   of the Lagrangian (\ref{matter}). First, it is a local enforcement as  the Ward Identities require that $\theta$ parameter and $\eta'$ field enter the effective Lagrangian in a specific way as eq. (\ref{matter}) states. Secondly,  it is a global enforcement as  the  $2\pi$ periodicity in $\theta$ implies there existence of interpolating $(\chi, \eta')$ configurations which inevitably present in the system. 

 We emphasize once again that the long range structure revealed in this section  might be the trace of a similar structure measured on the lattices, 
 as the transition from  high temperature weak coupling regime to low temperature   strong coupling  regime  should be smooth without any phase transitions on the way \cite{Yaffe:2008}. However, similar to our 
 comment in section \ref{dw},  we expect  that
  the $(\chi, \eta')$ domain walls determined by eqs. (\ref{set}) with boundary conditions (\ref{bc}) and asymptotical behaviour 
    (\ref{eta3}, \ref{eta5}) become very crumpled and wrinkled  objects with large number of foldings. Such fluctuations obviously increase the entropy of the DWs which eventually may overcome the intrinsic tension as holographic picture suggests, see \cite{Zhitnitsky:2011aa} and references therein. In fact, 
  it is quite likely that an appropriate description for this physics should be formulated in terms of holographic dual model, however we leave this subject for future studies.

 \section{Chiral Magnetic Effect (CME) and other topological phenomena}\label{CME}
 
 As we already mentioned,    the  ultimate goal of the present work is to understand  the  infrared physics  in the presence of external magnetic field, which is precisely the environment relevant for study of the CME. 
  We are now in position to couple our system (\ref{matter}) to external  Maxwell $U(1)$ field $A_{\mu}$ as massless quark $\psi$   carries the electric charge $e$ of  $A_{\mu}$ field.  In this section we use conventional Minkowski metric   in order to  compare 
  the obtained below formulae with known expressions   written normally  in Minkowski space-time. The $\eta'$ field which appears in  the low energy Lagrangian  (\ref{matter}) does not couple   to electromagnetic field directly as it is a neutral field.
  However, it does couple via triangle anomaly, similar to the textbook example describing $\pi^0\rightarrow 2\gamma$ decay. 
  The corresponding Maxwell  term $ S_{\gamma}$     and anomalous term $S_{\eta'\gamma\gamma}$   have the form
    \be
  \label{gamma}
   &S&_{\gamma} = -\frac{1}{4}\int d^{4}x \;  \left[ F_{\mu\nu}^2  \right]. \nonumber \\
  &S&_{\eta'\gamma\gamma} =  \frac{e^2N}{16 \pi^{2}} \int  d^{4}x \;  \left[  \eta' F_{\mu\nu} \tilde{F}^{\mu\nu}\right].
 \ee
 The  structure  of the $S_{\eta'\gamma\gamma}$ is unambiguously fixed by the anomaly.  It  describes  the interaction of the Maxwell field\footnote{not to be confused with gluon field from eq. (\ref{topologicalcharge}), (\ref{thetaincluded})} $F_{\mu\nu}$ with matter field. 
 
 The interaction (\ref{gamma}) is normally used to describe $\eta'\rightarrow 2\gamma$ decay. However, in the context of the present work
 we treat $\eta'(x)$ as external background field describing the $\eta'$ DW discussed in previous section \ref{DW_eta}.
 Therefore, a number of new and unusual coherent effects will emerge as a result of long range structure represented by the $\eta'$ domain wall described above. Let us emphasize again that this long range structure being represented by the $\eta'$ component  of the $(\eta', \chi)$ domain wall  is the Euclidean long range configuration describing the tunnelling processes in Minkowski space -time, rather than a physical configuration in real Minkowski space-time.   Nevertheless, this $\eta'$ coherent component does interact  with $F_{\mu\nu}$ field as eq. (\ref{gamma}) dictates. 
 
 We start by rewriting the action (\ref{gamma}) using conventional vector notations for electric $ \vec{E}$ and magnetic  $\vec{B}$
 fields, 
 \be
 \label{total}
 S_{\gamma}+S_{\eta'\gamma\gamma}= \int d^{4}x \left[\frac{1}{2} \vec{E}^2 - \frac{1}{2} \vec{B}^2 -
  \frac{Ne^2}{4 \pi^2} \eta'   \vec{E}\cdot \vec{B}\right].~~~
\ee
One can immediately  see that in the electric field will be induced in the  presence of external magnetic field $\vec{B}_{\rm ext}$
in the extended region where $\eta'$ is not vanishing, i.e.
 \be
 \label{E}
    \vec{E}= \frac{Ne^2}{4 \pi^2} \eta'  \vec{B}_{\rm ext}, 
\ee
 which is precisely the starting formula in \cite{Kharzeev:2007tn} if we identify the coherent $\eta'$ component from the $(\eta', \chi)$ domain wall with induced $\theta$ parameter introduced in \cite{Kharzeev:2007tn}.  Furthermore, is we assume that the DW is  extended  
 along $(x,y)$ directions,   we get
 \be
 \label{E1}
 (\mathbb{L}_x \mathbb {L}_y) E_z=   \left( \frac{e N l}{2\pi}\right)   \eta'(z)
  \ee
  where   $\mathbb L$  is size of the system, and   integer number $l$ 
 is the magnetic flux of the system, $ \int d^2x_{\perp}
 {B^z_{\rm ext}}=\Phi/e= 2\pi l /e$.
Formula (\ref{E1})  identically coincides with eq. (7) from  \cite{Kharzeev:2007tn} if the induced $\theta$ parameter from that work is identified with  extended along $(x,y)$ configuration represented by the $\eta' (z)$ domain wall described in section \ref{DW_eta}. The induced electric field along $z$ obviously implies that the current will flow and charge will be separated along ${B}^z_{\rm ext}$.

Anomalous coupling (\ref{total}) also implies that there will be induced current as a result of coordinate dependence of the $\eta'$ field, 
\be
\label{J1}
J^{\nu}= -\frac{Ne^2}{8 \pi^2} \partial_{\mu}\left(\eta' \tilde{F}^{\mu\nu}\right), 
\ee
which is convenient to represent in vector notations as follows
\be
\label{J}
J_0=\frac{Ne^2}{4 \pi^2} \vec{\nabla} \eta'   \cdot\vec{B}_{\rm ext}~~, ~~~~~ \vec{J}=\frac{Ne^2}{4 \pi^2}\dot{\eta'}  \vec{B}_{\rm ext}.
\ee
where we assume that the external magnetic field $\vec{B}_{\rm ext}$ is coordinate independent. 
Formula (\ref{J1})  for the anomalous current has been   studied previously in literature in many fields, including particle physics, cosmology, condensed matter physics. In particular, in the context  
of the axion domain wall it was extensively discussed in \cite{Huang:1985tt}. 

In the present context relevant for the CME formula (\ref{J}) reduces to the well known result when the induced $\theta (x, t)$ parameter 
is identified with extended $\eta'$ domain wall from section \ref{DW_eta}. Indeed, 
integrating $\int J_0d^3x$ leads precisely to the known expression for the charge separation effect (CSE) along $z$,  
\be
\label{Q1}
Q=\int d^3xJ_0=\frac{Ne^2}{4 \pi^2} \int dz \frac{ d\eta'}{dz} \int d^2x_{\perp} B^z_{\rm ext} = Nle,~~~
\ee
where we took into account the boundary conditions for the $\eta'$ field (\ref{bc}) and replaced
 $ \int d^2x_{\perp} B^z_{\rm ext} =\Phi/e= 2\pi l /e $. Furthermore, the expression for $ \vec{J}$ in eq. (\ref{J}) can be presented in much more familiar way if one replaces $\dot{\eta'}\rightarrow \dot{\theta}\equiv 2\mu_5$ as our identification suggests. With these replacements  the expressions (\ref{J}) and (\ref{Q1}) assume their conventional forms which are normally used in  CME studies. One should comment here that while our solution for the $\eta'$ domain wall considered in section \ref{DW_eta} is time independent, in fact it actually describes tunnelling effects in strong coupling regime.  Therefore,  it  obviously becomes a time dependent configuration  with a typical time scale $\dot{\eta'}\sim \Lqcd$ in a course of a smooth transition from weak coupling to strong coupling regime, as discussed at the end of section  \ref{DW_eta}. However, one can not use very large magnitude for $\dot{\eta'}\sim\mu_5\sim 1$ GeV (as many people do) for numerical  estimates
 as the effective Lagrangian approach which leads to formulae (\ref{J}), (\ref{Q1}) can only be justified for small values $|\dot{\eta'}|\ll \Lqcd$, and marginally justified for $|\dot{\eta'}|\simeq \Lqcd$. For large $|\dot{\eta'}|\gg \Lqcd$ the effective Lagrangian approach can not be justified, and computations should be based on a different technique    when underlying QCD degrees of freedom, quarks and gluons (rather than  effective $\eta'$ field) play the dynamical role.

 One should say that there are many other interesting topological effects originated from similar anomalous terms as originally discussed  in 
    terms of hadronic fields in \cite{Kharzeev:2007tn,Son:2004tq}, and re-derived in terms of microscopical     quark fields in \cite{Metlitski:2005pr,Fukushima:2008xe}. In particular, if a system with chemical potential $\mu$   rotates with angular velocity $\vec{\Omega}$, there will be a current flowing along $\vec{\Omega}$. The  charges will be also separated along the same direction. To get corresponding formulae one should replace $e\vec{B}\rightarrow  2\mu\vec{\Omega}$ in eq. (\ref{J})  as discussed in \cite{Kharzeev:2007tn}, i.e.
    \be
\label{J2}
J_0=\frac{Ne\mu}{2 \pi^2} \vec{\nabla} \eta'   \cdot\vec{\Omega}_{\rm ext}~~, ~~~~~ \vec{J}=\frac{Ne\mu}{2 \pi^2}\dot{\eta'}  \vec{\Omega}_{\rm ext}, 
\ee
 which  precisely coincides with eq. (A2) from \cite{Kharzeev:2007tn} if one identifies $\eta'$ field with induced parameter $\theta$ from  \cite{Kharzeev:2007tn} as we already discussed above. Specific consequences of   effect  (\ref{J2})  relevant for heavy ion collisions were discussed quite recently in \cite{Kharzeev:2010gr} where the effect was coined as the chiral vortical effect (CVE). 
 
The main point of this section is as follows.  
 The long range structure discussed  in section \ref{DW_eta}   might be the trace of a similar extended structure measured on the lattices, 
 as the transition from  high temperature weak coupling regime to low temperature   strong coupling  regime  should be smooth \cite{Yaffe:2008}. It is important that this long range structure describes the tunnelling effects and represented by $(\chi, \eta')$ fields in deformed QCD.
  These configurations are not real physical configurations in Minkowski space-time. Nevertheless, these long range configurations do interact with real physical $E\&M$ field as a result of anomaly (\ref{total}). 
  
  Such an interaction transfers unphysical long range correlations (expressed in terms of the Euclidean configurations describing the tunnelling processes)  to physical long range correlated  $E\&M$ effects (\ref{E}, \ref{E1}, \ref{J}, \ref{J2}).  The corresponding coherent effects  are accumulated  on large  scales $\sim  \mathbb L$   where the boundary conditions for different topological sectors are imposed.  
  
  This toy model explicitly shows that the large observed intensity of the effect as studied  at RHIC and ALICE at the LHC~\cite{Voloshin:2004vk,Abelev:2009tx,collaboration:2011sma,Selyuzhenkov:2011xq,Abelev:2012pa} might be  due to  a coherent phenomena when the large observable asymmetry  is  a result of accumulation of a small effect over large distances  $\sim  \mathbb L\gg \Lqcd^{-1}$. In different words,  the CSE  given by formulae (\ref{Q1}),  CME given by formula (\ref{J}) and CVE given by formula (\ref{J2})  lead to a large magnitude for an asymmetry   in spite of the fact that  parameters $|\dot{\eta'}|\sim \mu_5\sim \mathbb L^{-1} $ remain parametrically small during entire tunnelling transition. 
  It should be contrasted with some other numerical estimates  when the large   intensity of the effect  is achieved  by choosing a  relatively  large  $  |\mu_5|\sim 1~ {\rm GeV} \gg \Lqcd\sim 0.1$ GeV  in which case there is no region of validity  for  the effective Lagrangian approach. 
    
   \section{Conclusion and future directions}\label{conclusion} 

The question which is  addressed
in the present work is  as follows: what is the physics
behind of the long range order which was  postulated  in  \cite{Kharzeev:2007tn}, and which is     apparently     a required element for CME and CVE to be operational. 
We attempt to answer this question using the ``deformed QCD" as a toy model where all computations are under complete theoretical control as
this model is a weakly coupled gauge theory.  Still, this model  has all the relevant crucial elements allowing us to study difficult and nontrivial questions which are known to be present in real strongly coupled QCD.  The study of these effects in this toy model reveals that the long range structure may result from  the tunnelling effects and represented by $(\chi, \eta')$ fields in deformed QCD. 

Apparently, such  kind of transitions with long range structure in strong coupling regime are happening all the time, as it is observed in the lattice simulations. One should expect that the corresponding configurations at strong coupling should be  very crumpled and wrinkled objects with large number of foldings
  in contrast with our smooth $(\chi, \eta')$ domain walls described in section \ref{DW_eta}. Such local fluctuations are expected to occur as  it provides a large entropy for these configurations to overcome their intrinsic tension.  
  If the entropy of the configurations is sufficiently large, the corresponding objects will have effectively vanishing tension, and   an arbitrary number of such objects can be formed and they can percolate through the vacuum, forming a kind of a vacuum condensate.
  Nevertheless, the crucial element of these DWs, the long range coherence, is not lost in transition from weak to strong coupling regime. 
  Precisely this feature, we believe,  is the key element why the observed asymmetries are sufficiently strong and not washed out as a result of a conventional short range QCD fluctuations with a typical size $\Lqcd$. 
  
  We suspect that all other conventional mechanisms based on e.g.  instanton/sphaleron transitions can not provide sufficient intensities observed at RHIC and the LHC as the observed asymmetries must be accumulated on large  scales of order $\sim  \mathbb L$ rather than on scales of order $\Lqcd^{-1}$. 
  
On phenomenological side, 
  the very basic observed features, such as energy  and charge independence, of measured asymmetries in heavy ion collisions are automatically and naturally satisfied within the  framework 
  based on long range order, see  recent papers  \cite{Zhitnitsky:2010zx, Zhitnitsky:2012im}. The same framework based on the idea of a coherent accumulation of the effect also provides a natural  explanation for a strong dependence on centrality
as observed at RHIC and ALICE at the LHC~\cite{Voloshin:2004vk,Abelev:2009tx,collaboration:2011sma,Selyuzhenkov:2011xq,Abelev:2012pa}.

  Essentially, our study in a simplified version of QCD provides   a precise and very specific realization of an old idea \cite{Buckley:1999mv,Buckley:2000aa} (see also \cite{Kharzeev:1998kz,Kharzeev:1999cz} where similar idea  was formulated in different terms),  that a macroscopically large domain  with $\theta_{ind}\neq 0$ can be formed in heavy ion collisions.
  Now we can precisely  identify this domain  characterized by $\theta_{ind}\neq 0$ with interpolating 
  long range $\eta'$ field which traces a pure glue configuration describing the  transition   between different topological sectors.
  In different words,  the  domains with   $\theta_{ind}\neq 0$ should  not be thought as   real extended  regions formed in Minkowski space-time as a result of collision. Rather it   should be understood as    long range Euclidean coherent configurations which saturate the tunnelling transitions in path integral computations. Nevertheless, these long range structure formulated in terms of auxiliary Euclidean configurations can be translated into observable long range effects (\ref{E}, \ref{E1}, \ref{J}, \ref{J2}) in Minkowski space as a result of anomalous coupling with physical  $E\&M$ field.

We should also mention that CME has been extensively studied in the lattice simulations~\cite{ Buividovich:2009wi,Buividovich:2009my,Buividovich:2010tn,Blum:2009fh}. Independently, very different lattice studies   reveal that the crucial topological configurations saturating the path integral are represented by  extended, locally low-dimensional sheets of topological charge embedded in 4d space~\cite{Horvath:2003yj,Horvath:2005rv,Horvath:2005cv,Alexandru:2005bn,Ilgenfritz:2007xu,Ilgenfritz:2008ia,Bruckmann:2011ve,Kovalenko:2004xm,Buividovich:2011cv}. Our analysis based on computations  in weakly coupled ``deformed QCD" suggests that the long range configurations which are responsible for CSE/CME/CVE effects are precisely the same objects   which we identify with long range extended objects from refs. \cite{Horvath:2003yj,Horvath:2005rv,Horvath:2005cv,Alexandru:2005bn,Ilgenfritz:2007xu,Ilgenfritz:2008ia,Bruckmann:2011ve,Kovalenko:2004xm,Buividovich:2011cv}.
We presented a number of arguments suggesting that this relation is in fact quite generic as it is based on topological features of the theory, rather than on a specific details of the model. Therefore, we conjecture that this  relation continues to hold in strongly coupled QCD. This conjecture can be explicitly tested in the  lattice simulations as essentially this conjecture suggests that the topological charge distribution as it is done in ~\cite{Horvath:2003yj,Horvath:2005rv,Horvath:2005cv,Alexandru:2005bn,Ilgenfritz:2007xu,Ilgenfritz:2008ia,Bruckmann:2011ve,Kovalenko:2004xm,Buividovich:2011cv} and   electric charge distribution in the presence of the background magnetic field are strongly correlated and follow each other.  Such a correlation also provides a new, and much easier  way to study the original topological charge distribution by putting the system into the background magnetic field and studying the electric charge distribution in ``quenched approximation"  as it is previously done in refs.~\cite{ Buividovich:2009wi,Buividovich:2009my,Buividovich:2010tn,Blum:2009fh}. 

Our final comment is as follows. The transition from weakly  coupled ``deformed QCD" to strongly coupled regime   should be  smooth.  Still, this transition is  beyond the analytical control within QFT framework. What would be  an appropriate tool to study this physics in strongly coupled regime? It is very possible that the description in terms of the holographic dual model may provide the required tools and technique. In fact, the long range structure is obviously present  in holographic  model as 
one can see from  computations of  the so-called ``Topological Casimir Effect" \cite{Zhitnitsky:2011aa, BKYZ} when no massless degrees of freedom are present in the system, but dependence of physical observables  exhibit  a power like sensitivity  to  size of the system $\sim  \mathbb L^{-p}$. This scaling is in huge contrast  with  $\sim  \exp(-\mathbb L )$ dependence which is normally expected  in  a conventional theory with  a gap. Some analogies  presented in \cite{Zhitnitsky:2011aa} are actually suggesting that the ground state of QCD behaves very similarly to some condensed matter systems which are known to lie in  topological phases.  The last word whether these analogies can be extended to the strongly coupled four dimensional QCD remains, of course,   the prerogative of the direct lattice computations.

 \section*{Acknowledgements}
  I am    thankful to  Gokce~Basar, Dima Kharzeev, Ho-Ung Yee,  Edward Shuryak  and other  participants of 
 the workshop ``P-and CP-odd effects in hot and dense matter", Brookhaven, June, 2012, for useful and stimulating discussions related to the subject of the present work. I am also thankful to Misha Polikarpov for useful comments   on  feasibility to study the  correlations between configurations relevant for CME as studied in 
 ~\cite{ Buividovich:2009wi,Buividovich:2009my,Buividovich:2010tn,Blum:2009fh}
 and topological charge distribution  as studied in \cite{Horvath:2003yj,Horvath:2005rv,Horvath:2005cv,Alexandru:2005bn,Ilgenfritz:2007xu,Ilgenfritz:2008ia,Bruckmann:2011ve,Kovalenko:2004xm,Buividovich:2011cv}. 
 This research was supported in part by the Natural Sciences and Engineering Research Council of Canada.

\end{document}